\documentclass{mem}
\usepackage{natbib}\usepackage{txfonts}\usepackage{balance}
\usepackage{graphicx}
\usepackage[a4paper]{hyperref}
\idline{000}{1}

\def \sw {{\it Swift}}

\def \hcm {\hbox {\ifmmode $ atom cm$^{-2}\else atom cm$^{-2}$\fi}}

\begin{document}

\title{\emph{Swift}: the science across the rainbow}
   \subtitle{Mission Overview and Highlights of Results}

  \author{P. Romano\inst{1}  }
   \offprints{P.\ Romano}
   \institute{INAF, Istituto di Astrofisica Spaziale e Fisica Cosmica, 
	Via U.\ La Malfa 153, I-90146 Palermo, Italy 
        \email{romano@ifc.inaf.it}      
             }

\authorrunning{P.\ Romano}

\titlerunning{\emph{Swift}: Mission Overview and Highlights}

\abstract{
I present an overview of the \sw\ mission, which was launched on November 20, 2004 to
discover and observe the most energetic of astrophysical phenomena, $\gamma$--ray bursts (GRBs). 
After almost 6 years in space the Observatory is in excellent shape, with all systems and instruments 
performing nominally and in burst chasing mode for an average of 97\,\% of the time. 
\sw\ is also a multi-purpose multi-frequency mission with the observing time evolving 
from mostly GRB targets, to mainly secondary science ones such as supernovae, 
cataclysmic variables and novae, active galactic nuclei, Galactic transients, active stars and 
comets. I present the most recent science highlights.  

}
\maketitle{}

        \section{Introduction:  Mission Overview} 

The \sw\ Gamma-Ray Burst Explorer \citep{Gehrels04:SWIFT} is a 
MIDEX NASA mission, whose instruments were built
with the participation of the United Kingdom and Italy, that 
was successfully launched on 2004 Nov.\ 20.  
It is a first-of-its-kind autonomous rapid-slewing mission to 
discover and observe the most energetic of astrophysical phenomena, 
$\gamma$--ray bursts (GRBs), and pioneers the way for future rapid-reaction 
and multi-wavelength missions, such as {\it Fermi}. The nominal lifetime was of 
two years, but has subsequently been extended in 2006, 2008, and 2010. 

The scientific goals of the mission are to determine the origin of GRBs,  
to classify them and search for new types, to determine how the GRB blast-wave evolves 
and interacts with the surroundings, to use GRBs to study the early universe, and  
to perform a sensitive hard X-ray survey. 

\begin{figure}[]
\resizebox{\hsize}{!}{\includegraphics[clip=true]{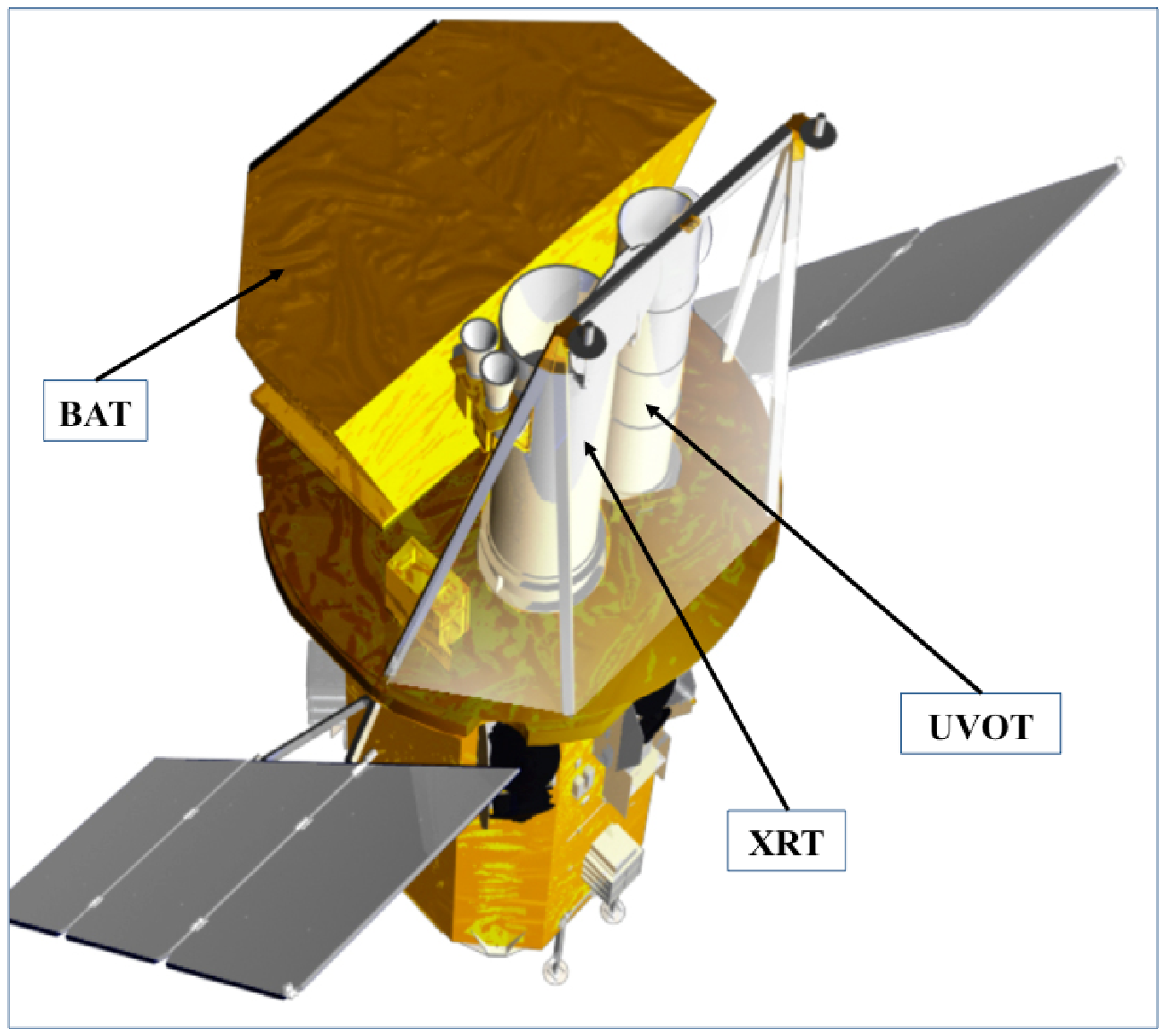}}
\caption{
\footnotesize
The  \sw\ satellite. 
}
\label{sait10:fig:swift}
\end{figure}

In order to tackle these goals, {\it Swift} carries three co-aligned telescopes (see Fig.~\ref{sait10:fig:swift}):
one wide-field instrument, the gamma-ray Burst Alert Telescope \citep[BAT, 15-150\,keV, ][]{Barthelmy2005:BAT}, 
which is the coded aperture mask telescope that provides the initial GRB trigger, 
and two narrow-field instruments (NFI), the X--ray Telescope  \citep[XRT, 0.2--10\,keV, ][]{Burrows2005:XRT},
and the UV/Optical Telescope \citep[UVOT, 1700--6500\AA, ][]{Roming2005:UVOT}. 
The main characteristics of the payload are reported in Table~\ref{sait10:tab:swift} 
\citep[see][ for a complete description]{Gehrels04:SWIFT}.

\begin{table}[t] 
\caption{\sw\ in-flight performance.}
\label{sait10:tab:swift}
\begin{center}
\begin{tabular}{lll}
\hline
BAT &  &  \\
    &FOV                   & 1.4 sr ( half-coded )\\
    &Energy range          & 15--150 keV \\
    &Position precision    & 2'--3'\\
\hline
XRT &  &  \\
    &FOV                   & 23.6'$\times$23.6'\\
    &Energy range          & 0.2--10  keV \\
    &Position precision    & 5''\\
\hline
UVOT &  &  \\
    &FOV                   & 17'$\times$17'\\
    &Energy range          & 1700--6500\AA \\
    &Position precision    & 0.5' \\
\hline
\end{tabular}
\end{center}
\end{table}

Much of the mission's success is undoubtedly due to the GRB observing strategy. 
The BAT detects a GRB and calculates the source location down to 2'--3', 
depending on the source brightness, and triggers an autonomous slew of the 
whole Observatory, so that the NFIs can image the GRB location, generally within
a minute or so. The BAT position is sent to the ground within 15\,s of the localization, 
so that the ground-based facilities can start observing the transient at once
and attempt to determine its redshift. 
Subsequently, the XRT can improve the GRB localization down to $\sim 5$'', while the UVOT
can further refine it to $\sim 0.5$''. In about 5 minutes, \sw\ can generally provide a
$\sim 0.5$'' position for the GRB, and  allows the investigation of the initial phases
of the afterglow starting as early as one minute after the BAT trigger. 
All this happens during a time span comparable to the one it takes to read this last paragraph.

        \subsection{Status of the Mission}

\begin{figure}[t]
\resizebox{\hsize}{!}{\includegraphics[clip=true,angle=270]{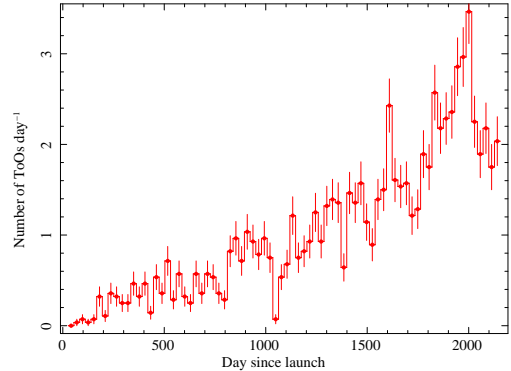}}
\caption{
\footnotesize
ToO observation requests as a function of time, on a monthly basis, since the start of the mission.
Courtesy of J.A.\ Kennea.   
}
\label{sait10:fig:toos}
\end{figure}

The \sw\ mission is in excellent shape, with an expected orbital life of more than 15 years. 
After almost 6 years (as of 2010-09-30) in space all systems are performing nominally 
with a science uptime of 97.5\,\%.   
We performed about 180,000 slews and have responded to the more than 500 bursts 
within an hour since the onset of the event with GCN circulars, so that the 
Flight Operation Team and Science Operation Team response has been excellent. 
Both the ground station and Observatory are in nominal condition. 
The same applies for the three instruments, BAT, XRT, and UVOT.   
BAT is burst chasing for 97\,\%  of the time, with a GRB discovery rate consistent 
with pre-flight predictions;  
XRT detects about 97\,\% of afterglows with a prompt response. 

\sw\ is also a multi-purpose multi-frequency mission whose observing time is 
evolving from an initial (as of 2006) 60\,\% of the time dedicated to GRB observations, 
and 30\,\% divided between target of opportunity (ToO, 10\,\%) 
and Fill-in observations (18\,\%), 
to a current 27\,\% dedicated to GRBs, 29\,\% to ToOs, and 26\,\% to Fill-in observations (2010). 
Figure~\ref{sait10:fig:toos} shows the number of ToOs received per day on a monthly basis
since the start of the mission, which is an ever-increasing function of time.

        \section{Science Highlights: GRBs}

\begin{figure}[t]
\resizebox{\hsize}{!}{\includegraphics[clip=true]{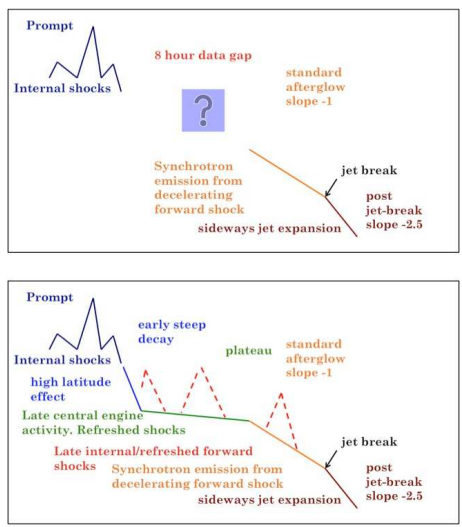}}
\caption{
\footnotesize
GRB light curves before (top) and after (bottom) \sw. 
}
\label{sait10:fig:stdlcvs}
\end{figure}

\sw\ has opened a new avenue of research in GRBs as, for the first time, 
we can study them in a continuous way, from prompt emission to all phases of 
the afterglow (AG). This is exemplified in Fig.~\ref{sait10:fig:stdlcvs} (top): 
at the times of {\it BeppoSAX} it used to take about 6--10 hours from the detection of 
a new GRB to the moment it could be imaged with lower-energy narrow-field instruments.
As the AG decays quite fast, this implied that at that point, 
the AG was a few orders of magnitude fainter, hence its localization more difficult to obtain,
and its study at lower energies much more complicated to perform.

\begin{figure*}[t!]
\resizebox{\hsize}{!}{\includegraphics[clip=true]{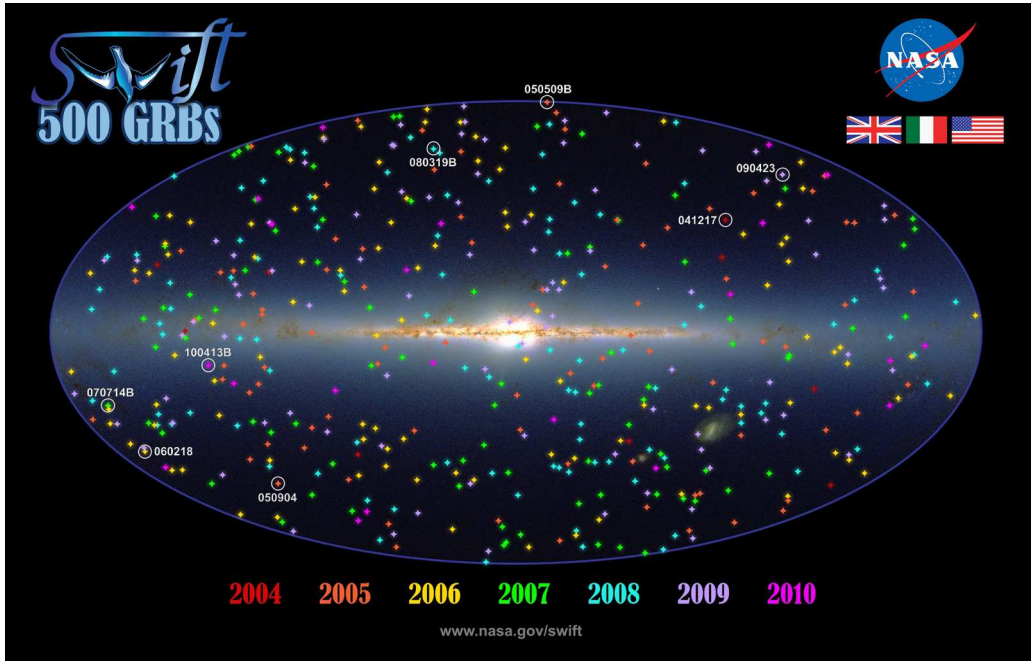}}
\caption{
\footnotesize
The first 500 GRBs observed by \sw. 
Credit: NASA/Goddard Space Flight Center/\sw.  
}
\label{sait10:fig:500}
\end{figure*}

\sw\ has now observed more than 500 GRBs (Fig.~\ref{sait10:fig:500}) panchromatically. 
At the date of writing (2010-09-30) \sw\ has detected 544 GRBs, 
84\,\% of which have an XRT-detected AG (this percentage rises to 97\,\% 
if the slew was performed immediately after the BAT trigger),
57\,\% with optical detection. 
Nowadays, about 88\,\% of the total number of AGs is due to \sw.  
Of all \sw\ GRBs, 174 have a redshift determination 
when only 41 existed before.  

These data have yielded a few surprises, the first being the fact that X--ray AGs 
do not follow the initial prompt decay to smoothly join the `standard' X--ray AG. 
On the contrary, the X--ray AG light curve has now a well-defined `canonical' shape, 
which can be divided in distinct phases 
(noted above the curve in Fig.~\ref{sait10:fig:stdlcvs}, bottom) for which 
several responsible mechanisms are invoked (reported below the curve). 
This `canonical'  \citep[][]{Zhang2006:thory_from_xrt,Nousek2006:lcvs,Obrien2006:xrtbat}
light curve is characterized by a steep/flat/steep/steeper shape, of which only the latter 
two were known before the launch of \sw. Many light curves also show flares 
\citep{Burrows2005:flarescience,Romano2006:050406,Falcone2006:050502b,Chincarini2007:flaresI}
that imply central engine long-lasting activity.  
A full review of the \sw\ results on GRBs can be found in \citet{Gehrels2009ARAA}. 
Here I shall mention only a few.

GRB~080319B ($z=0.937$) has been dubbed `the naked-eye burst' 
\citep[][and references therein]{Racusin2008:080319B}  
since it reached the uniquely bright visual magnitude of 5.3 mag, 
offered the brightest optical and  X--ray fluxes, 
and had one of highest $\gamma$--ray fluences ever recorded for a GRB
($\sim 6.3\times10^{-4}$ erg cm$^{-2}$, 20\,keV--7\,MeV). 
It has proven to be a powerful diagnostic of the
detailed physics of this explosion within seconds of its formation. 

GRB~090423, with its redshift of 8.2 \citep{Salvaterra2009:090423}, is
not only the most distant GRB, but also the most distant object ever observed
in the universe, and it has allowed us to establish that the 
the mechanisms and progenitors that gave rise to this burst about 600
million years after the Big Bang are not markedly different from those producing GRBs
10 billion years later.
The previous record-holders were 
GRB~050904 \citep[$z=6.3$][]{Cusumano2006:050904Nature}, 
GRB~080913 \citep[$z=6.7$][]{Greiner2009:080913}.

\begin{figure*}[t!]
\resizebox{\hsize}{!}{\includegraphics[clip=true]{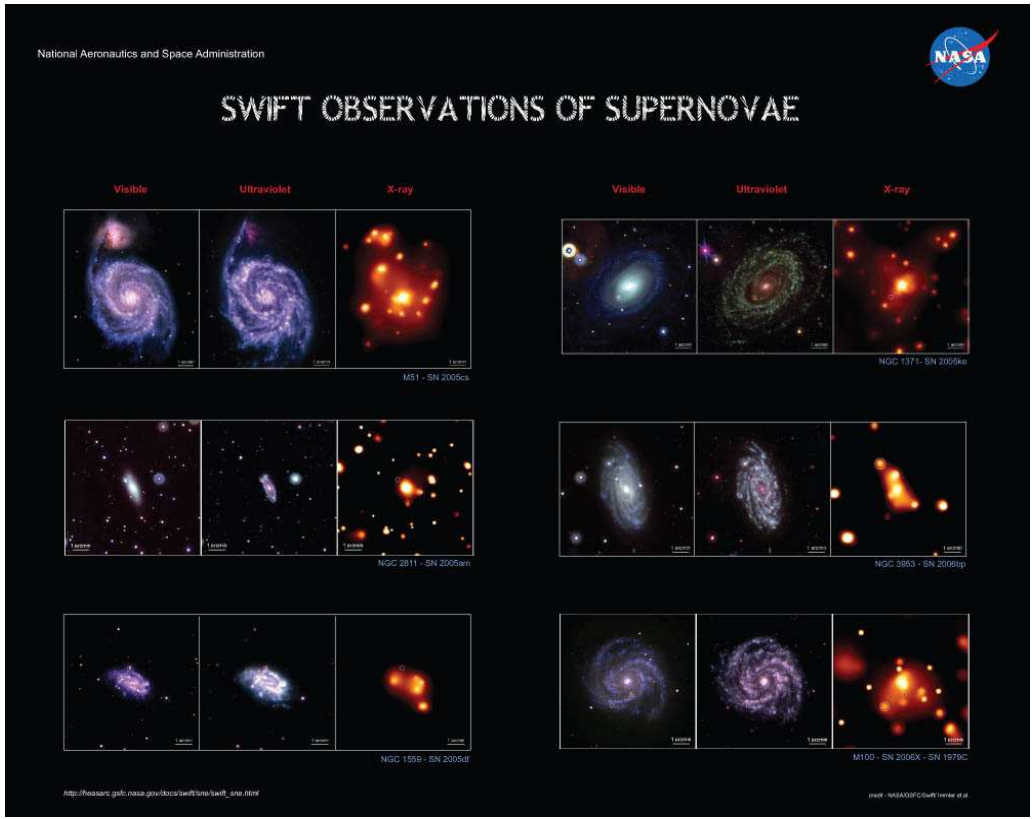}}
\caption{
\footnotesize
\sw\ optical (left), ultraviolet (middle), and X-ray images (right) of nearby SN host galaxies. 
The visible ($V$, $B$, and $U$ filters) and ultraviolet ($UVW1$, $UVM2$ $UVW2$ filters) images 
were obtained with UVOT. 
The X-ray images (0.2--10 keV) were obtained simultaneously with XRT. 
Credit: NASA/\sw/S.\ Immler. 
}
\label{sait10:fig:SNe}
\end{figure*}

\sw\ has detected 46 short GRBs, 70\,\% of which localized to arcsecond precision 
\citep[none were available before \sw, see, e.g.\ ][]{Gehrels2005:050509B}, 
and about half with host identification and/or redshifts. 
The main implication is that the progenitors are likely binary mergers, as opposed to 
the death of massive stars, which are the likely progenitors of long GRBs.   
Short GRBs show X--ray light curves similar to the ones of long GRBs, including the
presence of flares. 
 
The link between long GRBs and supernovae (SN) dates back before the \sw\ era,
with a firm association established between an almost-simultaneous Type Ib/Ic SN, 
so that long GRBs can be the high-energy counterpart of such SN events. 
However, it was only with GRB~060218 that we could 
actually observe the beginning of a SN explosion and its intimate link with a GRB. 
\citet[][]{Campana2006:060218} describe how, in addition to the classical non-thermal emission, 
GRB~060218 showed a thermal component in its X-ray spectrum, 
which cooled and shifted into the optical/UV band as time passed, and 
interpreted these features as arising from the break-out of a shock wave driven 
by a mildly relativistic shell into the dense wind surrounding the progenitor,
which was probably a Wolf-Rayet star. 

\begin{figure*}[t!]
\resizebox{\hsize}{!}{\includegraphics[clip=true]{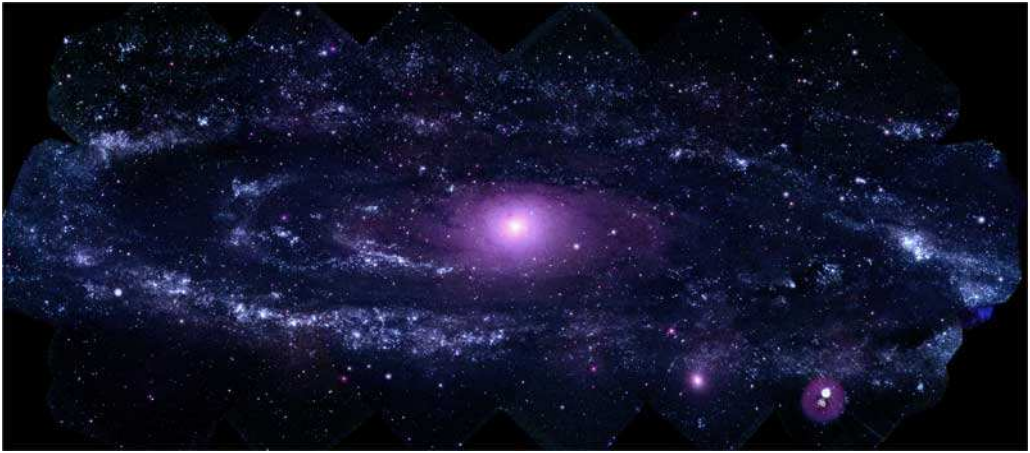}}
\caption{
\footnotesize
The highest-resolution ultraviolet image of the Andromeda Galaxy obtained by 
\sw. The image is a mosaic of 330 images acquired between May 25 and July 26 2008, 
for a total exposure time of 24 hours, by UVOT.
UVOT revealed about 20,000 ultraviolet sources in a region 200,000 light-years wide. 
Several features are apparent in the new mosaic like the striking difference between 
the galaxy's central bulge and its spiral arms. 
Credit: NASA/\sw/Stefan Immler (GSFC) and Erin Grand (UMCP). 
}
\label{sait10:fig:andromeda}
\end{figure*}

\sw\ has been studying SNe extensively (see Fig.~\ref{sait10:fig:SNe}). 
In particular, \citet[][]{Brown2009:sne_uvot} present more that 100 SNe light curves 
in the UV and X--rays, more than those obtained with all previous missions combined. 
This large sample has allowed characterization of light curve shapes and the classification of SNe based on
UV-optical photometry and can also be compared with rest-frame UV observations of high-redshift 
SNe observed at optical wavelengths.
\sw\ is also surveying nearby galaxies like M~31 (Fig.~\ref{sait10:fig:andromeda}) 
in order to understand the star-formation conditions 
and relate them to the  conditions in the distant galaxies where GRB are occurring.

        \section{Science Highlights: secondary science}

\sw's multi-wavelength capability and flexible observing schedule
make it perfectly suited for studies of comets, cataclysmic variables (CVs) and novae, 
active stars, active galactic nuclei (AGN), and Galactic transients. 
The increasing number of ToO requests is an indication that the astronomical community has 
realized this and is exploiting the potentiality of this Observatory for non-GRB science.

Among the most interesting objects that \sw\ has observed we can find several comets. 
These objects are uncontaminated fossils from the early days of the Solar system, 
  hence precious tools in studying cosmogony. 
Comets are difficult to observe in that they are fast-moving targets, hence requiring a fast
accurate pointing, and are often too bright for {\it XMM-Newton} and {\it Chandra}. 
\sw\ has allowed to study the composition  (with the UVOT grisms), the chemistry (UVOT grisms), 
the evolution of
the gas surrounding the comet (UVOT imaging), and the interaction of this gas with the 
Solar wind (XRT). Some of the main results can be found in \citet[][]{Bodewits2010HEAD}. 

A true renaissance has occurred in the field of accreting white dwarfs and novae
because of the availability of simultaneous optical to hard X-ray observations. 
\sw\ has made unique observations, collecting whole-outburst light curves showing 
the rise and fall of the super-soft emission. 
As an example, it has been possible to study the rare outburst of the dwarf nova GW Lib 
\citep[][]{Byckling2009:GWLib}, and to find that its long inter-outburst time-scale with 
no observed normal outbursts is consistent with the idea that the inner disc is evacuated 
or the disc viscosity is very low. 
On the other hand, \citet[][]{Evans2009:GKPer} show that the large disc in the 
dwarf nova-like GK Per is able to maintain a long-term memory of the mass transfer rate
from the secondary.

\begin{figure}[]
\resizebox{\hsize}{!}{\includegraphics[clip=true]{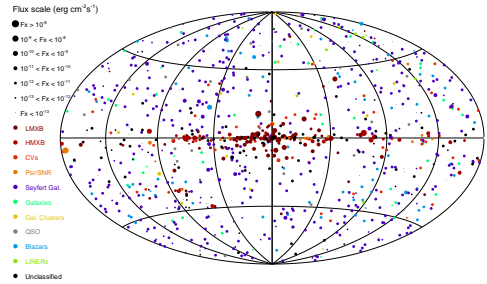}}
\caption{
\footnotesize
Map (Galactic coordinates) of the sources detected and associated with a 
counterpart, during the first 54 months of survey performed by BAT. 
Different colors indicate different kinds of sources, while the 
symbol dimension is proportional to the 15-150\,keV flux of the source. 
Adapted from \citet{cusumano2010:54mo}. 
}
\label{sait10:fig:batsur}
\end{figure}

\sw's contribution to the AGN field is considerable, since it covers the 
critical optical/UV to hard X-ray range of the spectral energy distribution (SED) 
where the transition between the synchrotron and inverse Compton emission usually occurs. 
AGN are notoriously highly variable sources in the UV as well as in X-rays, 
so the requirement is simultaneity of such observations. 
This allows researchers to determine the physical conditions in individual AGNs,
and to treat AGN as a class, studying their evolution and formation. Furthermore,
\sw\ has allowed to study particle acceleration mechanisms in blazar jets. 
As an example, \citet[][]{Abdo2010:SEDFermiblazars} study the SED of Fermi bright blazars, 
and propose a new classification scheme based on the position of the peaks of the SED. 
%
In Fig.~\ref{sait10:fig:batsur} I show the results based on a 54-month survey of the sky
\citep[][]{cusumano2010:54mo}. The BAT, because of its large field of view and detector area, 
offers the opportunity to significantly increase the number of detections contributing
to the luminosity of the sky in the hard X-rays allowing a substantial 
improvement of our knowledge of the AGN and of the cosmic hard X-ray background.

\begin{figure*}[ht!]
\resizebox{\hsize}{!}{\includegraphics[clip=true]{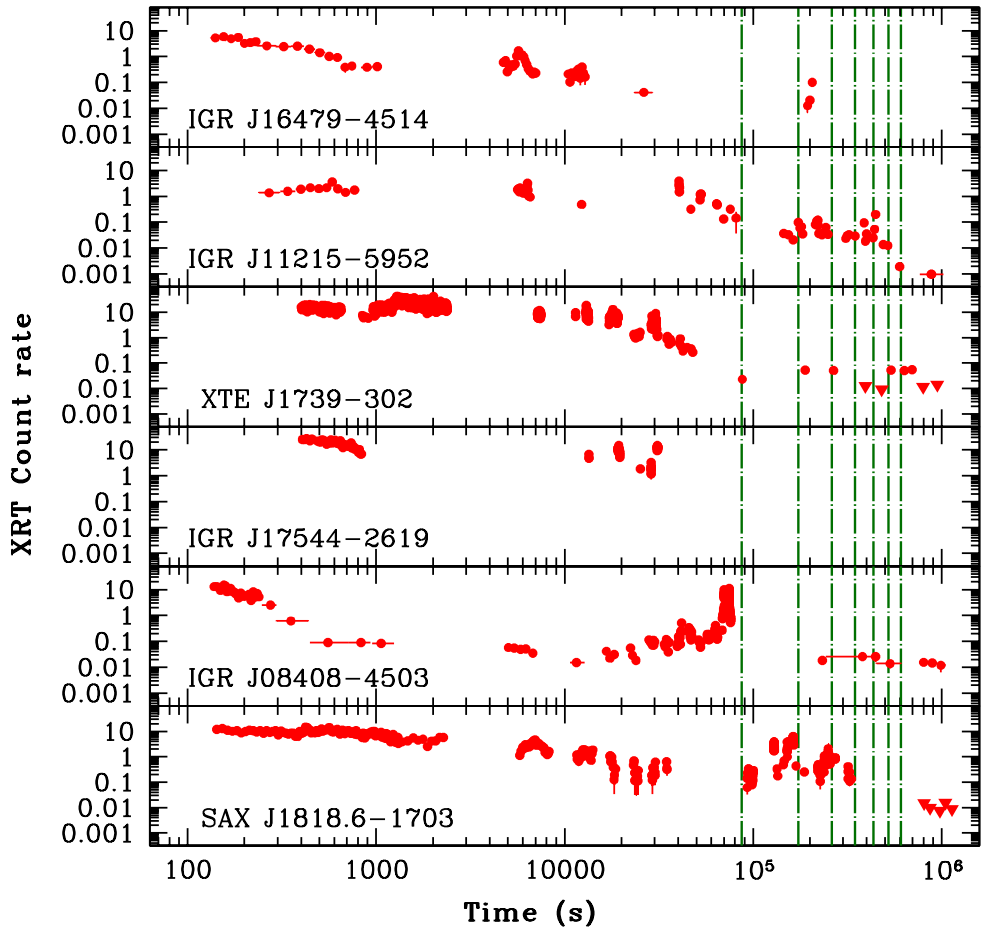}}
\caption{
\footnotesize 
Light curves of the outbursts of SFXTs followed by Swift/XRT
referred to their respective triggers. 
Points denote detections, triangles 3$\sigma$ upper limits.
Note that where no data are plotted, no data were collected.  
Vertical dashed lines mark time intervals equal to 1 day, up to a week. 
}
\label{sait10:fig:sfxts}
\end{figure*}

Finally, a mention should be made to Galactic transients, 
among them the Very Faint X--Ray Transients 
and Supergiant Fast X--Ray Transients.
The latter are an ideal test-case that shows the potentials of \sw\ in the field of transients.
On one hand \sw\ is the only observatory which
can detect their outbursts (see Fig.~\ref{sait10:fig:sfxts}) 
in their very early stages and study them 
panchromatically as they evolve, as is done for GRBs. 
On the other hand, \sw's flexible observing scheduling makes a monitoring 
cost-effective, so that \sw{} has given SFXTs the first non serendipitous 
attention through monitoring campaigns that cover all phases of their lives 
with a high sensitivity in the soft X--ray regime, where most SFXTs had not been observed
before 
\citep[see ][and references therein]{Romano2010:sfxts_paperVI}.

        \section{Conclusions and future perspectives}

\sw\ offers a unique combination of autonomous fast repointing and flexible scheduling which,
combined with the wide energy range, has made it a successful mission for GRB studies.
The same properties, in addition to the support given to other missions such as {\it INTEGRAL}, 
{\it AGILE}, and {\it Fermi}
have also made it the `go-to' facility for non-GRB targets,
such as CVs and novae, AGNs, and Galactic transients, as testified by the fact that 
about 55\,\% of the \sw\ publications are nowadays on secondary science topics.

\begin{acknowledgements}
I would like to thank the conference organizers for their kind invitation and for providing a pleasant, colourful 
meeting environment. I thank the whole {\it Swift} Team 
starting from the PI, N.\ Gehrels, for running such a tight, enthusiastic ship,
the instrument leads, S.\ Barthelmy, D.N.\ Burrows, and P.\ Roming, the Italian PI, G.\ Chincarini, 
and especially the science planners and duty scientists, for their professional juggling 
of a multitude of conflicting requests without losing their sense of humour.  
This work was supported by contracts ASI I/011/07/0. 
\end{acknowledgements}


\end{document}